\documentclass[12pt]{article} 
\usepackage{a4wide,latexsym}
\usepackage{amsmath}
\usepackage{array}
\usepackage{bbm}
\usepackage{cite}
\usepackage{accents}

\newcommand\lsim{\mathrel{\rlap{\lower4pt\hbox{\hskip1pt$\sim$}}
    \raise1pt\hbox{$<$}}}
\newcommand\gsim{\mathrel{\rlap{\lower4pt\hbox{\hskip1pt$\sim$}}
    \raise1pt\hbox{$>$}}}

\normalsize  
\begin{document}
\parskip=3pt plus 1pt

\begin{titlepage}
\vskip 1cm
\begin{flushright}
\today \\
\end{flushright}

\setcounter{footnote}{0}
\renewcommand{\thefootnote}{\fnsymbol{footnote}}

\vspace*{1.5cm}
\begin{center}
{\Large\bf James Chadwick: ahead of his time} 
\\[25mm]

{\normalsize\bf Gerhard Ecker}\\[1.2cm]
University of Vienna, Faculty of Physics \\[10pt]
Boltzmanngasse 5, A-1090 Wien, Austria \\[10pt]
\end{center}

\vspace*{2cm}

\begin{abstract}
\vspace*{.2cm} 
\noindent
James Chadwick is known for his discovery of the neutron. Many of his
earlier findings and ideas in the context of weak and strong nuclear
forces are much less known. This biographical sketch attempts to
highlight the achievements of a scientist who paved the way for
contemporary subatomic physics.

\end{abstract}

\vfill

\setcounter{footnote}{0}
\renewcommand{\thefootnote}{\arabic{footnote}} 

\end{titlepage}

\newpage

\section{Early years} 
\label{sec:early}
\renewcommand{\theequation}{\arabic{section}.\arabic{equation}}
\setcounter{equation}{0}

James Chadwick was born on Oct. 20, 1891 in Bollington, Cheshire in
the northwest of England, as the eldest son of John Joseph Chadwick
and his wife  Anne Mary. His father was a cotton
spinner while his mother worked as a domestic servant.  In 1895 the
parents  left Bollington to seek a better life in Manchester. James
was left behind in the care of his grandparents, a parallel
with his famous predecessor Isaac Newton who also grew up with his
grandmother. It might be an interesting topic for sociologists of
science to find out whether there is a  
correlation between children educated by their grandmothers and future
scientific geniuses. 

James attended Bollington Cross School. He was
very attached to his grandmother, much less to his parents.
Nevertheless, he joined his parents in Manchester around 1902
but found it difficult to adjust to the new environment. The family
felt they could not afford to send James to Manchester Grammar
School although he had been offered a scholarship. Instead, he attended
the less prestigious Central Grammar School where the teaching was
actually very good, as Chadwick later emphasised. 

Chadwick's modest background did not prevent him from receiving an excellent
general education. Especially his mathematics teacher encouraged him
and finally persuaded him to enter a competition for a scholarship at 
Manchester University, which he won at the age of sixteen\cite{brown:1997}.

\vspace*{.2cm} 

\section{University studies} 
\label{sec:univ}
\renewcommand{\theequation}{\arabic{section}.\arabic{equation}}
\setcounter{equation}{0}

In May 1907, Ernest Rutherford moved from McGill University in Canada
to Manchester University. He inherited from his predecessor a modern
and well-equipped department with one major deficiency: there was no
radium. The problem was solved by Stefan Meyer from the Radium
Institute in Vienna with a generous gift of some 300 milligrams of
radium chloride. The Manchester School of radioactivity under
Rutherford's leadership soon produced results that would
revolutionise science. 

In the fall of 1908, Chadwick came to the university for a
preliminary interview. Although he intended to take up mathematics, he
ended up being interviewed by a lecturer from the physics
department. Chadwick was too shy to admit his mistake and thus started
his life as a physicist by accident. He was not too impressed by his
first-year courses but things changed substantially when in his 
second year he attended lectures on electricity and magnetism
delivered by Rutherford, ``the first stimulating lectures I had ever had in
physics'' \cite{weiner1}. 

At the end of the second year, there was no more formal education in
physics. Instead, the few remaining honours students in physics were
assigned specific research projects by Rutherford. As Chadwick later
remarked \cite{weiner1}: ``I had half an education in physics. There
were whole aspects of physics I knew little about.'' 

At that time,
there were no generally accepted units for radioactivity. At an
international congress in Brussels in September 1910 it was agreed
that the amount of radioactivity released by a gram of radium would
serve as the standard unit, later to be known as a curie. The
third-year research project assigned to Chadwick was a highly
topical one. He was instructed to investigate a method initially devised by
Rutherford to compare different radium sources. When Chadwick became
aware of a small problem in Rutherford's 
original setup he did not dare to mention it to the master, risking
rather to disappoint Rutherford who noticed the problem after the
first measurements. The approach was applied to a comparison of the two
standard radium sources available at the time, one initially provided
by Marie Curie and the other by Stefan Meyer. The comparison was
complety convincing, leading at one hand to the acceptance of an agreed
world standard and, on the other hand, to Chadwick's first published
paper together with Rutherford\cite{cr:1912}.

In the summer of 1911, Chadwick graduated with first class honours
although the final exam was by no means straightforward. The written part
presented no problems. On the other hand, Chadwick found out only
shortly before the exam that he also had to undergo a practical
examination, with J.J. Thomson as external examiner. Chadwick claims
that he was terrified by Thomson's charisma and could hardly say
a word\cite{weiner1}. In any case, Rutherford was convinced of
Chadwick's talents 
and accepted him as a graduate student. As a demonstrator he received
his first salary, modest enough but sufficient for regular lunches 
after three years as an impoverished undergraduate. The growing impact
of the Manchester School attracted many visitors, among them Niels
Bohr who came in March 1912 as a postdoctoral fellow and made friends
with Chadwick. The most prominent result of Bohr's stay was his atomic
model, which dominated the physicists' view of the atom till the
emergence of quantum mechanics and beyond. 

The next project of Chadwick was a thorough investigation of
$\gamma$-ray absorption in gases. Until that time all information on
$\gamma$-ray absorption coefficients had been indirect. Using simple but
ingenious techniques, Chadwick was able to measure these coefficients
with uncertainties of only a few percent. This research definitely
established Chadwick as an imaginative experimenter. Moreover, 
in addition to obtaining precise results he was able to draw concise
and far-reaching conclusions in the ensuing publications. In the
present  case, his first paper as sole author \cite{ch:1912}, he
pointed out that the high concentration of ions found in the upper 
atmosphere could not be wholly due to radiation from the
radioactive material in the earth and, therefore, must be the result of
radiation from outer space. In the same year 1912 the
Austrian physicist Viktor Hess established the existence of cosmic
rays after ionization measurements on a series of seven balloon
flights \cite{hess:1912}. 

Chadwick received his M.Sc. in the summer of 1912. In the following
year, the university nominated him for a prestigious 1851 Exhibition
Science Research Scholarship. Rutherford wanted Chadwick to spend at
least one year of  this scholarship in his group in Manchester but the
terms of the scholarship foresaw that the recipient would have to work
at an institute other than the nominating university. Despite Rutherford's
interventions, the scholarship commissioners remained adamant: take it
or leave it. It was finally agreed that Chadwick would spend the first
year of tenure either in Berlin with Rutherford's former collaborator
Hans Geiger or at the University of Vienna. No one, least the
scholarship commissioners, could foresee that Chadwick was about to
embark on one of his life's great adventures\cite{brown:1997}.

\vspace*{.2cm} 

\section{First World War}  
\label{sec:FWW}
\renewcommand{\theequation}{\arabic{section}.\arabic{equation}}
\setcounter{equation}{0}

In the fall of 1913, Chadwick arrived in Berlin to spend the
first year of his stipend in Geiger's laboratory at the
Physikalisch-Technische Reichsanstalt in Charlottenburg, a suburb of
Berlin. Geiger had returned from Manchester to Germany in 1912 after
performing seminal experiments with his student Ernest Marsden that
established Rutherford's atomic model. He gave Chadwick a warm
welcome and introduced him to other colleagues such as Otto Hahn and
Lise Meitner from the Kaiser Wilhelm Institute for Chemistry. After quickly 
obtaining a working knowledge of German and getting used to German
bureaucracy\cite{weiner1}, Chadwick soon enjoyed the friendly
atmosphere and Geiger's helpful guidance. His chosen area of work was
$\beta$ radiation. He could not anticipate that this topic would
remain a crucial and much debated subject in the development of
subatomic physics for at least another twenty years. Nor could he
anticipate that his contribution, although completely correct, would
remain contested for almost fifteen years.

\subsection{Beta decay}
\label{subsec:beta}
During the first decade of the 20$^{\rm th}$ century it was established that 
$\alpha$ particles are helium nuclei and $\beta$ particles are
electrons. It was also found that in a given $\alpha$ decay the
emitted $\alpha$ particles all had the same energy corresponding to
the mass difference between initial and final nuclei. It was natural
to assume the same behaviour for the emitted electrons in
$\beta$ decay. But by 1913, researchers from both the Manchester School and
Hahn's laboratory had discarded the hypothesis that the
$\beta$ spectrum was monochromatic. Instead, the experiments seemed to 
show that $\beta$ spectra consisted of  several discrete lines. To that
date, the weak point of all experiments was the detection of the emitted
electrons on photographic plates. When Chadwick entered the game he
had the advantage of employing instead of a photographic plate a
particle counter that had just been developed by Geiger and that has
been carrying his name ever since. After initial doubts about his results he
convinced himself that the line spectrum was actually an artefact of
the photographic detection. In the publication \cite{ch:1914}, a
hallmark of modern physics that is unfortunately difficult to access
(the original publication was probably translated into German by
Geiger\cite{brown:1997}), he made it absolutely clear that the accepted picture of
Rutherford, Hahn and others was incorrect and that the
$\beta$ spectrum was actually continuous. While Rutherford immediately
accepted Chadwick's results, many others like Meitner remained
sceptical. 

After the war, Rutherford suggested to Charles Ellis, a young member
of the Manchester group (see also Subsec.~\ref{subsec:ruhleben}),
to reexamine the issue of the $\beta$ spectrum. Ellis not only
confirmed Chadwick's findings of 1914 but he also explained the
occurrence of discrete lines superimposed on the continuous spectrum
as being due to internal conversion involving electrons in the atomic
shell\cite{ellis:1921}. Lise Meitner was not convinced and
insisted\cite{meitner:1922} that the primary electrons in $\beta$
decay must all have the 
same energy because of energy conservation. Although
her reasoning was theoretically correct she could not explain the
continuous spectrum found by Chadwick and Ellis blaming it on
problems with their experimental setups. Since Chadwick's original
result was called into question by Meitner he teamed up with Ellis for
a measurement of the intensity distributions of electrons in the
$\beta$ decays of $_{~82}^{214}Pb$ (radium B) and $_{~83}^{214}Bi$
(radium C) by an ionisation method. They summarised their results as
follows\cite{chell:1922}: ``Firstly, the continuous spectrum has a real
existence which is not dependent on the experimental arrangement and
any explanation of it as due to secondary causes is untenable \ldots'' Due
to her theoretical ``prejudice'', Meitner was still not convinced.  It took
another five years of hard work by Ellis and collaborators before the
matter was finally settled. In 1927, Ellis and Wooster undertook an
absolute measurement \cite{ellwoo:1927}  of the heat produced by the
total absorption of  the electrons emitted in the $\beta$ decay of
$_{~83}^{210}Bi$. They demonstrated that the average energy
released per individual $\beta$ decay  was equal to the mean energy of
the continuous spectrum and that secondary processes, as called for by
Meitner, did not exist. In a follow-up experiment, Meitner and her
colleague Orthmann not only confirmed \cite{meiort:1930} the results of 
Ellis and Wooster but they arrived at an even stronger
conclusion. Ellis and Wooster had speculated that some continuous
$\gamma$ spectra could save energy conservation because 
$\gamma$ rays could not be observed in their calorimeter. But  Meitner
and Orthmann showed employing special counters that such a continuous
$\gamma$-ray spectrum did not exist. 

The experimental situation was now settled but the theoretical
dilemma became even worse.  As the results appeared to contradict the
sacrosanct conservation of  energy, Bohr speculated\footnote{Actually,
  Bohr upheld his view at least until 1932.} at the end of the
1920s that in the microcosm energy conservation might
only hold on average, but that an individual decay
could violate the energy balance. But there was also a problem with
the conservation of angular momentum if the electron with its spin 1/2
were the only decay product in addition to the final nucleus. 
Moreover, there were also problems with quantum statistics. According
to the general picture of the nucleus at the time, the nucleus of the
nitrogen isotope $^{14}_{~7}\!N$ should contain fourteen protons and seven
electrons. Because of the odd number of particles with spin 1/2 the
nitrogen nucleus would have half-integer spin and should 
satisfy Fermi-Dirac statistics. But experiment actually showed that the
$^{14}_{~7}\!N$ nucleus had integer spin and was therefore a boson.
Finally, it was difficult to reconcile with quantum mechanics and in
particular with the uncertainty relation that particles as light as
electrons could be confined in such a small volume as an atomic
nucleus. 

In December 1930, Pauli broke the Gordian knot in his famous
letter to the  
``radioactive ladies and gentlemen'' who gathered for a meeting in
T\"ubingen. He proposed as a solution of the various problems that the
electron in $\beta$ decay is accompanied by an additional particle
that would have to be electrically neutral and have spin 1/2. Pauli
named the postulated particle neutron but soon the name neutrino (the
small neutron) proposed by Enrico Fermi was generally accepted (see
also Sec.~\ref{sec:neutron}). In the
presence of this neutrino, the conservation of both energy and angular
momentum would be restored. The mass difference between initial and
final nuclei  determines the total energy of the decay products. This
energy is now shared between electron and neutrino, hence the
continuous energy spectrum of the electron. Because the neutrino has
spin 1/2 the conservation of angular momentum is also
guaranteed. Pauli wrote in his letter that he did not dare to
publish his idea for the time being. Proposing a particle that could
possibly never be detected experimentally was too daring at that
time. Three years later, after the discovery of the
neutron by Chadwick (Sec.~\ref{sec:neutron}), Fermi formulated the 
first quantum field theory of $\beta$ decay \cite{fermi:1934}.

\subsection{Particle-wave duality}
Back in 1914, Chadwick started a project scattering $\beta$ particles
on a thin gold foil, in analogy to the famous experiments of Geiger
and Marsden with $\alpha$ particles. In the course of that work he
suggested to Geiger \cite{weiner1} ``that perhaps electrons might be
scattered from a crystal surface in much the same way as
X rays. Geiger said there was nothing in it, it was rather a silly
suggestion to make.''  Nine years later, a certain Louis de Broglie
had the same idea \cite{broglie}, inspiring in particular Erwin
Schr\"odinger to set up his wave mechanics. In this case, the relevant
experiments were indeed performed \cite{davisson,reid}, demonstrating
wave-particle duality\footnote{Wave-particle duality was already well
  established for photons at that time.} also for matter particles. Once 
again, Chadwick was ahead of his time but the following events would
have prevented him anyway from actively pursuing his idea.

\subsection{Science in an internment camp}
\label{subsec:ruhleben}

Chadwick's career came to an abrupt end in August 1914 when the First
World War broke out. After the invasion of Belgium by the German army
Great Britain declared war on Germany and the situation became
precarious for a British citizen in Berlin. Although his German
colleagues in the laboratory were very helpful as Chadwick later
acknowledged gratefully he was arrested in November 1914 as an enemy
alien. Together with hundreds of other British civilians, he was
interned in a camp (Engl\"anderlager) in Ruhleben near
Spandau west of Berlin. Initial hopes that the war would be over by
Christmas soon faded and four years full of privation followed. 

This period is described in much detail in Brown's biography
\cite{brown:1997}. Here, I confine myself to the remarkable social
activities in the camp, in particular in the form of an Arts and
Science Union where Chadwick played a prominent role. For instance, he
delivered regular lectures on electricity and magnetism and even on
radioactivity that were well received by the participants. One of them
was Charles Ellis who as a cadet of the Royal Military
Academy had the bad luck of spending his summer holidays together with
colleagues in Germany at the outbreak of the war. All of them were
interned in Ruhleben.

Ellis developed a strong interest in science and soon became
Chadwick's favourite student. After the war, the two men continued to
work together in Cambridge where Ellis became a world authority for the
physics of $\beta$ and $\gamma$ rays (see
Subsec.~\ref{subsec:beta}). While in the camp, Chadwick managed to
organise 
a small laboratory where they performed several experiments
mainly in chemistry. He was even given for inspection some radioactive
toothpaste that became popular in Germany at that time. For getting
both equipment and scientific literature the camp authorities were
remarkably cooperative. Chadwick was even allowed to leave the camp
for visiting the prominent German scientists Heinrich Rubens, Walther
Nernst and Emil Warburg who all offered help. 

The general living conditions were less agreeable. By 1917 the
blockade of the British Navy on merchant ships had a devastating
effect on food supplies to Germany in general and to the inmates of the
Ruhleben camp in particular. Chadwick was severely undernourished and
had serious digestive problems that would accompany him for the rest of
his life. After the armistice in November 1918 all internees were
released. After a long journey via the Baltic and the North Sea, 
Chadwick finally arrived at his parents' home in Manchester.

\vspace*{.2cm} 

\section{Strong interaction} 
\label{sec:strong}
\renewcommand{\theequation}{\arabic{section}.\arabic{equation}}
\setcounter{equation}{0}

\vspace*{.2cm} 

As early as 1815, the English chemist 
William Prout suspected on the basis of existing measurements of
atomic masses that all atoms are built up of hydrogen atoms (Prout's
hypothesis). Scattering $\alpha$ particles on light atoms, in
particular on nitrogen, Rutherford demonstrated that the hydrogen
nucleus (denoted proton by him in 1920) does indeed occur in all
nuclei  \cite{rutherford:1919}. But Rutherford also recognised that 
additional, electrically neutral constituents must be contained in the 
nuclei in order to explain nuclear masses. He called these
constituents neutrons and he pictured them as bound states of protons
and electrons. Two reasons seemed to support such a picture. The mass
of these neutrons was comparable with the proton mass and the
negatively charged electrons would compensate the electrostatic
repulsion between the protons at least to some extent. 

Scattering $\alpha$ particles on protons, Rutherford found 
deviations from the Coulomb law (electrostatic repulsion between
$\alpha$ and $p$) \cite{rutherford:1919b}: ``\ldots not inconsistent
with the view that the forces between colliding atoms augment rapidly
for values of $d < 3.5 \cdot 10^{-13}$ cm.'' 
He associated the deviations with the complex 
nature of  $\alpha$ particles as bound states of four protons and
two electrons. With an improved experimental
setup, Chadwick repeated the experiment, first as part of his doctoral
thesis at Cambridge and then more thoroughly together with a young
colleague, the Swiss-born Canadian Etienne Bieler. Their results
confirmed Rutherford's findings but their conclusions went 
beyond. Investigating the differential cross section for various 
scattering angles, they observed agreement with expectations 
for slow $\alpha$ particles (low energies). On
the other hand, the measured numbers of scattered protons greatly
exceeded the expectations for fast $\alpha$ particles (high
energies), in one case 100 times as large, with only the Coulomb force
between point 
charges. In the latter case also the angular distribution of
the scattered protons was different. Their main conclusion 
\cite{cb:1921} was hailed by Abraham Pais  \cite{pais:1988} as marking 
the birth of the strong interactions: ``\ldots no system of four H
nuclei (i.e. protons) and two electrons united by inverse square law
forces could give a field of force of such intensity over so large an
extent \ldots It is our task to find some field of force which will
reproduce these effects \ldots The
present experiments do not seem to throw any light on the
nature or the law of variation of the forces at the seat of an
electric charge, but merely show that the forces are of very
great intensity.''   

During the following years, the Cambridge group extended their studies
by scattering $\alpha$ particles on heavier atoms, more specifically
on helium, magnesium and aluminium. These attempts were reviewed in the
classic monograph of Rutherford, Chadwick and Ellis in 1930
\cite{rce:1930}. The results for $\alpha \, \alpha$ scattering were
similar to those from $\alpha \, p$ scattering \cite{rc:1927}. Extending
the experiments to the heavier atoms magnesium and aluminium did not
clarify the situation. One reason was that the lever arm for
distinguishing between the Coulomb force and any additional force is
smaller for heavier atoms. From his results, Bieler \cite{bieler:1924}
concluded that the additional force was attractive and seemed to vary
with distance as $r^{-4}$ although a dependence as $r^{-3}$ could not
be ruled out either. On the other hand, Debye and Hardmeier
claimed that a force varying as $r^{-5}$ would also describe the data
by assuming that the incoming $\alpha$ particle strongly
polarises the heavy nuclei \cite{dh:1926}.
However, as emphasised in Ref.~\cite{rce:1930},  it did not seem
possible to explain the collisions  with hydrogen or helium nuclei in
the same way. 

At the end of the 1920s, the situation was as unclear as at the
beginning of the decade. Once again, Chadwick and his colleagues were
some fifteen years ahead of their time. For an understanding of the 
experimental results two fundamental theoretical developments were
necessary, the quantum mechanical scattering theory and the Yukawa
potential with the pion mass setting the scale for the onset of the new
force \cite{yukawa:1935}. 

From 1935 on, significant progress was made
in the understanding of the strong interactions of nucleons and
mesons. Mainly on the basis of nonrelativistic potential models
involving mesons,
it became possible to explain nuclear structure and nuclear
reactions. However, these achievements were restricted to reactions
where nucleons have small relative velocities. The developments
leading to the formulation of a relativistic quantum field theory of
the strong interactions (quantum chromodynamics) and its incorporation
in the Standard Model of the fundamental interactions can for instance
be found in Ref.~\cite{GE}.

\section{Discovery of the neutron} 
\label{sec:neutron}
\renewcommand{\theequation}{\arabic{section}.\arabic{equation}}
\setcounter{equation}{0}
Asked by Charles Weiner whether he had thought that his 
work done on the neutron might be of Nobel Prize calibre, Chadwick
answered\cite{weiner1} : ``The award of a Prize, it seems to me, to be
not so much a question of luck but a question of being there at the
right time.'' In 1932, Chadwick was indeed right on time. In
consequence, he was ennobled by the Nobel Foundation in 1935
and by the English King in 1945 where the second knighthood also
acknowledged his role in the Manhattan Project.

Chadwick's discovery of the neutron thus differed from
many of his 
earlier achievements where he happened to be ahead of his time. Since
the neutron discovery is described in much detail in many
books\cite{brown:1997} and articles\cite{wiki1}, I will confine myself
here to a brief summary of events for completeness. 

Rutherford had introduced the notion of a
neutron as bound state of proton and electron already in 1920 in
order to understand nuclear masses. Especially after the advent of
quantum theory, the difficulties of this picture became more and more
acute. As already mentioned in Subsec.~\ref{subsec:beta},
a serious discrepancy with experiment had to do with
quantum statistics. In Rutherford's model the nucleus of the
nitrogen isotope $^{14}_{~7}\!N$ contained fourteen protons and seven
electrons. Because of the odd number of particles with spin 1/2 the
nitrogen nucleus would have half-integer spin and should 
satisfy Fermi-Dirac statistics, contradicting experimental 
evidence\footnote{The same
  problem existed for the $^{6}_{3}\!Li$  nucleus.}.
In addition, while a bound state of proton and electron was
well understood in the form of the hydrogen atom, it was difficult to
reconcile with the uncertainty relation that particles as light as
electrons could be confined in such small volumes as atomic
nuclei.  

Already in 1930, Walther Bothe and Herbert Becker\cite{bb:1930} had
scattered energetic $\alpha$ particles from a polonium source on
several light elements such as $^{9}_{4}\!Be$. They observed a very
penetrating radiation that was not deflected by an electric field and
was therefore interpreted as $\gamma$ rays. Two years later, 
Ir\`{e}ne and Fr\'{e}d\'{e}ric Joliot-Curie repeated the
experiment\cite{jcj:1932}. When the unknown radiation was directed at
some hydrogen-containing material such as paraffin wax, it released
high-energy protons. Therefore, the process was interpreted as
proton Compton scattering. The problem was that this would have
required $\gamma$ rays with unrealistically high energy (50 MeV). The 
Italian theorist Ettore Majorana commented
sarcastically\cite{maj:1932}: ``What fools, they have discovered the
neutral proton and they do not recognize it.''

Neither Chadwick nor his mentor Rutherford believed the interpretation
as proton Compton scattering. Chadwick immediately set to work and
within three strenuous weeks not only repeated the French experiment 
but also scattered the radiation on various atoms other than
hydrogen. He found that 
the new radiation consisted not of $\gamma$ rays, but of uncharged
particles with about the same mass as the proton. The last sentence in
his Nature article\cite{ch:1932a} is crystal clear: ``Up to the present,
all the evidence is in favour of the neutron, while the quantum
hypothesis (i.e. the $\gamma$-ray hypothesis) can only be upheld if
the conservation of energy and momentum is relinquished at some
point.''   

One may ask the question why Chadwick was more successful than the
Joliot-Curies, both experienced scientists who would receive the Nobel
Prize in chemistry in 1935 for their discovery of artificial
radioactivity. A plausible answer is that Chadwick was not only right
on time but he was also in the right surroundings. While there was a
general consensus that atomic nuclei consisted of protons and
electrons, the idea of a neutron as bound state of proton and electron
was rather specific to the Cavendish Laboratory. This is for instance
spelled out by Joliot\cite{joliot} when commenting on Chadwick's
discovery: ``The word neutron had already been used by the
genius Rutherford in 1923 (actually 1920), at a conference, to denote
a hypothetical neutral particle which, together with protons, made up the
nucleus. This hypothesis had escaped the attention of most physicists,
including ourselves. But it was still present in the atmosphere of the
Cavendish Laboratory where Chadwick worked and it was natural -- and
just  -- that the final discovery of the neutron should have been made
there.'' Another indication is that Pauli in his letter of December
1930 to the ``radioactive ladies and gentlemen'' proposed to call his
hypothetical particle ``neutron'' as if he had never heard of
Rutherford's neutron. Fortunately, Fermi came to the rescue and
renamed Pauli's particle neutrino to avoid confusion.  

Whether the neutron is a bound state of proton and electron or not
was still an open question for some time. The neutron mass is a
crucial parameter to answer this question. If
\begin{equation} 
m_n ~~ > ~~ m_p + m_e ~,
\end{equation}  
the neutron cannot be a bound state and it would be as
elementary a 
particle as the proton. The precise measurement of the
neutron mass turned out to be very demanding.
Several measurements of
the neutron mass published by different  groups were 
not conclusive and allowed for both possibilities. 
Initially, Chadwick favoured the bound-state nature of the
neutron\cite{ch:1932b}: ``It is, of course, possible
to suppose that the neutron may be an elementary particle. This view has
little to recommend it at present, except the possibility of explaining the
statistics of such nuclei as $^{14}\!N$.''  

In 1933 Maurice Goldhaber, a young refugee from Nazi-Germany, was
offered a research position at the Cavendish by Rutherford. At some
point, he suggested to Chadwick that the deuteron (usually called
diplon at the time) might be a good candidate for a precise measurement
of the neutron mass if it could be split by bombarding it with
$\gamma$ rays:
\begin{equation}
\gamma +~  ^{2}_{1}\!H \rightarrow p + n ~.
\end{equation} 
Their experiment\cite{cg:1934} produced a precise value for the
neutron mass showing that the neutron cannot be a bound state. 
Electrons were banished from atomic nuclei, which consist of protons and
neutrons only.

\vspace*{.2cm} 

\section{Later years} 
\label{sec:epi}
\renewcommand{\theequation}{\arabic{section}.\arabic{equation}}
\setcounter{equation}{0}
In the fall of 1935, shortly before receiving the Nobel Prize,
Chadwick moved to Liverpool University. He refurbished the 
old-fashioned laboratories and initiated the construction of a
cyclotron, making Liverpool one of the European centers of nuclear 
physics. As leading expert on neutron physics, he was chosen to write
the final draft of the so-called MAUD Report, the basis of American-British
collaboration in the Manhattan Project. His presence at the Trinity
nuclear test was characterised by a science journalist associated to
the Manhattan Project\cite{laurence}: ``Never before in history had any
man lived to see his own discovery materialize itself with such telling
effect on the destiny of man.''
In 1948, Chadwick moved back to Cambridge to become Master of Gonville
and Caius College, where he had been a research student in the early
1920s. He retired at the end of 1958 and moved to North Wales with his
wife. Ten years later they once more moved back to Cambridge to be
near their daughters. James Chadwick died in his sleep on July 24,
1974.

\vspace*{.6cm} 

\paragraph{Acknowledgements}
I have made extensive use of the excellent biography of Andrew
Brown\cite{brown:1997} and of the interviews recorded by Charles
Weiner\cite{weiner1}. For more detailed studies of Chadwick's
biography, especially also of his personality, I very much recommend
these two references. For suggestions, corrections and other help,
I thank Walter Grimus, Helmut Neufeld, Maria Probst, Michael Springer
and Brigitte Strohmaier.


\vspace{.6cm}

\begin{thebibliography}{10}
\bibitem{brown:1997}  
A. Brown, {\it  The neutron and the bomb: a biography of Sir James
  Chadwick}, Oxford Univ. Press, New York, 1997.

\bibitem{weiner1}
C. Weiner, {\it Sir James Chadwick, oral history}, American Institute of
Physics, College Park, Maryland, 1969.

\bibitem{cr:1912}
E. Rutherford and J. Chadwick, Proc. Phys. Soc. 24
 (1912) 141.

\bibitem{ch:1912}
J. Chadwick, Proc. Phys. Soc. 24 (1912) 152.

\bibitem{hess:1912}
V.F. Hess, Phys. Zeits. 13 (1912) 1084.

\bibitem{ch:1914}
J. Chadwick, Verhandlungen der Deutschen Phys. Gesellschaft 16 
(1914) 383.

\bibitem{ellis:1921}
C.D. Ellis, Proc. Royal Soc. A99 (1921) 261.

\bibitem{meitner:1922}
L. Meitner, Zeits. Physik 9 (1922) 131.

\bibitem{chell:1922}
J. Chadwick and C.D. Ellis, Proc. Cambridge Phil.
Soc. 21 (1922) 274.

\bibitem{ellwoo:1927}
C.D. Ellis and W.A. Wooster, Proc.  Royal Soc. A117
(1927) 109.

\bibitem{meiort:1930} 
L. Meitner and W. Orthmann, Zeits. Physik 60 (1930) 143. 

\bibitem{fermi:1934}
E. Fermi, Zeits. Physik 88 (1934) 161.

\bibitem{broglie} L. de Broglie, Ann. de Physique 3 (1925) 22.

\bibitem{davisson}
C. Davisson and L.H. Germer, Nature 119 (1927) 558.

\bibitem{reid}
G.P. Thomson and A. Reid, Nature 119 (1927) 890.

\bibitem{rutherford:1919}
E. Rutherford, Phil. Mag., Ser. 6, Vol. 37 (1919) 581.

 \bibitem{rutherford:1919b}
E. Rutherford, Phil. Mag., Ser. 6, Vol. 37 (1919) 537.

\bibitem{cb:1921}
J. Chadwick and E.S. Bieler, Phil. Mag., Ser. 6, Vol. 42 (1921) 923.

\bibitem{pais:1988} 
A. Pais, {\it Inward bound: of matter and forces in the physical
  world}, Oxford Univ. Press, New York, 1986. 

\bibitem{rce:1930}
E. Rutherford, J. Chadwick and C.D. Ellis, {\it Radiations from
  radioactive substances}, Cambridge Univ. Press, Cambridge, 1930. 

\bibitem{rc:1927}
E. Rutherford and J. Chadwick, Phil. Mag., Ser. 7, Vol. 4 (1927) 605.

\bibitem{bieler:1924} 
E.S. Bieler, Proc. Royal Soc. A105 (1924) 434.

\bibitem{dh:1926}
P. Debye and W. Hardmeier, Phys. Zeits. 27 (1926) 196.

\bibitem{yukawa:1935}
H. Yukawa, Proc. Phys. Math. Soc. Japan 17 (1935) 48.

\bibitem{GE}
G. Ecker, {\it Particles, fields, quanta: from quantum mechanics to
  the Standard Model of particle physics}, Springer Nature Switzerland
AG, Cham, 2019.

\bibitem{wiki1} 
https://en.wikipedia.org/wiki/{Discovery\!\_of\_the\_neutron}

\bibitem{bb:1930} 
W. Bothe and H. Becker, Zeits. Physik 66 (1930) 289.

\bibitem{jcj:1932}
I. Joliot-Curie and F. Joliot, Comptes Rendus 194 (1932) 273.

\bibitem{maj:1932}
In E. Segr\`{e}, {\it From x-rays to quarks}, W.H. Freeman, New York, 1980.

\bibitem{ch:1932a} 
J. Chadwick, Nature 129 (1932) 312.

\bibitem{joliot}
In M. Goldsmith, {\it Fr\'{e}d\'{e}ric Joliot-Curie}, Lawrence and
Wishart, London, 1976.

\bibitem{ch:1932b} 
J. Chadwick, Proc. Royal Soc. A136 (1932) 692.

\bibitem{cg:1934}
J. Chadwick and M. Goldhaber, Nature 134 (1934) 237.

\bibitem{laurence}
W.L. Laurence, {\it Dawn over zero: the story of the atomic bomb},
A.A. Knopf, New York, 1946.

\end{thebibliography}
\end{document}